\documentclass[preprint,aps,tightenlines,floatfix,showpacs]{revtex4}
\usepackage{graphics}
\usepackage{bm}
\usepackage{amsmath}
\usepackage{amssymb}

\begin{document}

\title{Linking the exotic structure of ${}^{17}$C to its unbound mirror 
${}^{17}$Na}
  \author{K. Amos$^{(1)}$} 
  \email{amos@unimelb.edu.au}
  \author{L. Canton$^{(2)}$}
  \author{P. R. Fraser$^{(2)}$}
  \author{\mbox{S. Karataglidis$^{(3)}$}}
  \author{J. P. Svenne$^{(4)}$}
  \author{D. van der Knijff$^{(1)}$}

  \affiliation{$^{(1)}$ School  of Physics,  University of  Melbourne,
    Victoria 3010, Australia}
  \affiliation{$^{(2)}$ Istituto  Nazionale  di  Fisica  Nucleare,
    Sezione di Padova, Padova I-35131, Italia}
  \affiliation{$^{(3)}$ Department of Physics, University of Johannesburg, 
    P.O. Box 524 Auckland Park, 2006, South Africa}
  \affiliation{$^{(4)}$ Department  of  Physics  and Astronomy,
    University of Manitoba, and Winnipeg Institute for Theoretical Physics,
    Winnipeg, Manitoba, Canada R3T 2N2}

\date{\today}

\begin{abstract}
 The structure of ${}^{17}$C is used to define a nuclear
interaction that, when used in a multichannel algebraic scattering
theory for the $n$+${}^{16}$C system, gives a credible definition 
of the (compound) excitation spectra. When couplings to the
low-lying collective excitations of the ${}^{16}$C-core are taken into
account, both sub-threshold 
and resonant states about the $n$+${}^{16}$C threshold are found.
Adding Coulomb potentials to that nuclear
interaction, the method is used for the mirror system of $p$+${}^{16}$Ne
to specify the low-excitation spectrum of
the particle unstable $^{17}$Na. We compare the results with those
of a microscopic cluster model. 
A spectrum of low excitation resonant states in ${}^{17}$Na is found
with some differences to that given by the microscopic-cluster model.
The calculated resonance half-widths (for proton emission) range
from $\sim 2$ to $\sim 672$ keV. 
\end{abstract} 


\maketitle


\section{Introduction}

The  spectra of radioactive nuclei at or just 
beyond a drip line are most intriguing. To date, details of their spectra
are poorly known at best. Few if any excited states have been identified.
Likewise the spin-parities of many of the known states have not been or are 
uncertainly
assigned.  Nowadays, opportunities exist to investigate
spectra of such exotic systems using isotope separator on-line facilities
with which production of radioactive ion beams having energies
typically $0.1A$ to $10A$ MeV is possible. 
Reactions using these beams with higher
incident energy, particularly from hydrogen targets,  can be, and have been, 
used to study the structure of 
the radioactive ions as well~\cite{Ka06,Ka08}. 
However, it is the low energy domain that
interests us as we wish to consider structures of compound systems 
formed by amalgamation of the beam ion and a nucleon. 

With light mass systems having charge number $\pi$ and neutron number
$\nu$, there is often the possibility to link
the structures of mirror systems.  Usually there is  a reasonably well known 
spectrum of a nucleus (${}_{\pi = Z}^{A+1}X_{\nu = N+1}$), 
which we treat as
a compound of  a neutron ($n$) with 
${}_{\pi = Z}^{\hspace*{0.4cm}A}X_{\nu = N}$ to 
define a chosen nuclear model interaction.
With that model, assuming charge invariance
of the nuclear force and adjusting for Coulomb effects,
the spectrum of the mirror, ${}_{\pi = N+1}^{\hspace*{0.4cm}A+1}Q_{\nu = Z}$ 
may be predicted. 
This has been done~\cite{Ca06}, for example, for the
mass-7 isobars with multichannel algebraic scattering (MCAS)~\cite{Am03} 
evaluations of the spectra of the compound systems; 
${}^7$Li (as $n$+${}^6$Li), ${}^7$Be (as $p$+${}^6$Li), 
${}^7$He (as $n$+${}^6$He), and ${}^7$B (as $p$+${}^6$Be).
Also, the approach predicted a spectrum for the
(particle unstable) nucleus ${}^{15}$F when treated as a compound 
of $p$+${}^{14}$O~\cite{Ca06a}.
In that study, the nuclear interaction 
was set by an analysis of the mirror system ${}^{15}$C treated as 
$n$+${}^{14}$C.
It was found that key requirements for obtaining resonance states of 
${}^{15}$F were, a) the
Coulomb barrier, which is essential in recreating the resonance aspect
of the observed spin $\frac{1}{2}^+$ ground state,
 b) coupling of the extra core proton to 
distinct states of ${}^{14}$O and c) consideration of  the Pauli principle
within a coupled-channel collective model prescription.
The latter two features ensured a credible
sequence of spin-parity values, a very good fit to the high-quality
elastic scattering cross sections known at that time~\cite{Go04,Gu05},
and prediction of a set of narrow resonances only a few MeV
above the (two) known ones. Subsequently, narrow resonances in the region 
of that excitation energy were observed~\cite{Ca06a,Mu09,Mu10}.
(Note: There is an error in Ref.~\cite{Mu09} relating to
citation of the results of our earlier study~\cite{Ca06a}.
In their Table I,
references 6 and 7 have been reversed in both the table caption and
the header row. A similar error has also been made in Table I in  
Ref.~\cite{Mu10} involving their references 16 and 17.)

Recently,  Timofeyuk and Descouvemont~\cite{Ti10} used the same 
philosophy of fixing the nuclear aspect by a two-center, microscopic
cluster model (MCM) of ${}^{17}$C, treated as $n$+${}^{16}$C, to find 
a spectrum of ${}^{17}$Na as $p$+${}^{16}$Ne.  They predict 
narrow resonances in the low excitation spectrum of the particle unstable 
nucleus, ${}^{17}$Na, with very broad ones above that.
Those results were a stimulus to use the MCAS approach as a complementary  
study.
Thus we consider the mirror systems, ${}^{17}$C ($n$+${}^{16}$C)
and ${}^{17}$Na ($p$+${}^{16}$Ne), especially since 
the low excitation spectrum
of ${}^{17}$C has been found experimentally~\cite{El05,Bo07,Sa08} and
microscopic models for the structure of that nucleus have been 
proposed~\cite{El05,Bo07,Ka08,Ti10}. 
An MCAS analysis of the low excitation spectrum of
${}^{17}$C has been made before~\cite{Ka08}. 
However, the results of that study came from
an overly simple, two-channel, evaluation.  Therein, also, we 
considered a distorted wave approximation (DWA) analysis of inelastic 
scattering data~\cite{Sa08} of $70A$ MeV ${}^{17}$C ions from hydrogen targets. 
Those DWA evaluations were made using no-core shell model wave functions
and the complete set of results allowed us to pose some constraints upon the 
structure of ${}^{17}$C. 

However, inadequacies remained due, in part, to simplifications in the 
previous MCAS evaluations based upon limited available data. 
Subsequent experiments~\cite{Bo07} have suggested a number of 
additional spin-parity assignments.
The spin-parities of the three closely spaced sub-threshold 
states of ${}^{17}$C are identified and 
current no-core shell models fail to  match them adequately.
It has been suggested~\cite{Ti10} that 
coupling of a neutron to states at about 4 MeV excitation in
${}^{16}$C is needed to improve the results. 
That coupling was not included in our previous study~\cite{Ka08}
and so we present herein results of MCAS evaluations in which 
coupling to the $4^+$ state in the mass-16 nuclei is included.  

\section{MCAS evaluations of ${}^{17}$C as an $n$+${}^{16}$C system}

\begin{figure}[h]
\scalebox{0.8}{\includegraphics*{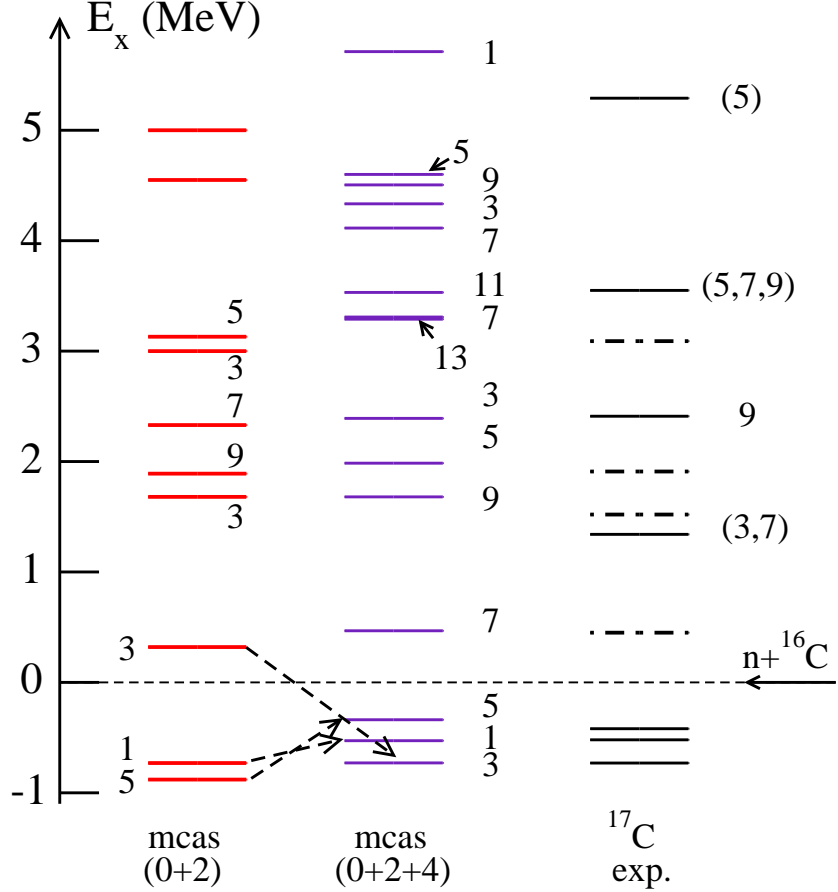}}
\caption{\label{C17spect}
MCAS results for the spectrum of ${}^{17}$C compared with experimental
data.
The energy scale is set with the $n$+${}^{16}$C threshold
as zero emphasising that the lowest three states are
stable against neutron emission.}
\end{figure}

MCAS calculations of the $n$+${}^{16}$C system were made
using three states in ${}^{16}$C; the $0^+$ (ground), $2^+$ (1.766 MeV),
and $4^+$ (4.142 MeV) states. We assume that couplings to the other
states (presumed $0^+_2, 2^+_2$, and $3^+$) in the spectrum between 3
and 5 MeV excitation are not strong and that the rotational 
model for the interactions, as defined previously~\cite{Am03}, suffices 
in seeking
the spectrum of ${}^{17}$C. The parameter values required to get
the results displayed in Fig.~\ref{C17spect} are listed in 
Table~\ref{params}.
\begin{table}[h]
\begin{ruledtabular}
\caption{\label{params}
The MCAS parameter values used to  define the channel-coupling 
properties of the $n$+${}^{16}$C system. Energy units are MeV,
length units are fm.}
\begin{tabular}{cccc}
$V_0$    & $V_{ll}$ &  $V_{ls}$ & $V_{Is}$ \\
-36.7 & -2.0 & 9.0 & 1.7 \\
\hline
R & a & $\beta_2$ & $\beta_4$\\
2.9 & 0.8 & 0.33 & 0.1\\
\hline
state in  & & OPP $\lambda_{lj}$ & \\
${}^{16}$C & ($1s_{\frac{1}{2}}$, $1p_{\frac{3}{2}}$,
$1p_{\frac{1}{2}}$) &      $1d_{\frac{5}{2}}$ & $2s_{\frac{1}{2}}$ \\       
\hline
$0^+$ ({\rm ground}) & 10$^6$ & 2.7 &  0.0\\
$2^+$ (1.766) & 10$^6$ & 2.7 &  0.0\\
$4^+$ (4.142) & 10$^6$ & 0.0 &  2.0\\
\end{tabular}
 \end{ruledtabular}
\end{table}
The OPP scale values coincide with Pauli blocking of the $1s$
and $1p$ orbits while the $1d$ and $2s$ values  equate to 
a Pauli hindrance in those orbits as was needed
in predicting~\cite{Ca06} narrow states in the spectrum of ${}^{15}$F; 
a nucleus beyond the proton drip line.

The potential parameter values and Pauli blocking/hindrance
weights are quite similar to the set used
previously~\cite{Ka08}, now with inclusion of a small hexadecapole
deformation to link the $4^+$ state to the ground state in first order
and with some Pauli hindrance of the $2s_{\frac{1}{2}}$ orbit 
in the connections to the $4^+$ target state.

The spectrum that results is shown in Fig.~\ref{C17spect} and
labelled ``mcas(0+2+4)". 
It is compared with the spectrum found
previously from the 2-state MCAS evaluation (``mcas(0+2)"), and with
the experimentally-known one that is a combination of states listed in
Table 4 of Ref.~\cite{Bo07} and in Fig.~5 of Ref.~\cite{Ra96}. The
energies of that tabled spectrum have been adjusted by $-0.729$~MeV;
the value of the $n$+${}^{16}$C threshold in ${}^{17}$C. Some states
specified by  Raimann {\it et al.}~\cite{Ra96} are displayed by the 
dash-dot lines. 
Only positive parity states are known in the low excitation spectrum, and the
integers associated with the individual energy levels shown are two
times their spin.

Clearly the three known subthreshold ($n$+${}^{16}$C) states
are now matched well in energy and spin-parity by the new MCAS
results. The other known and uncertain spin-parity states also
have matching MCAS partners in proximity of their excitation
energies.  Additionally, the uncertain states from Raimann 
{\it et al.}~\cite{Ra96} seem to have possible matches, and 
the first low-lying state above threshold of that set 
we expect to be a $\frac{7}{2}^+$
resonance. The (0+2+4) MCAS spectra have  a number of aspects in common with
that shown in Fig.~1  of Ref.~\cite{Ti10}. Besides the three closely 
spaced and weakly bound sub-threshold states, the MCM study gave a group 
of states ($\frac{3}{2}^+, \frac{5}{2}^+, \frac{7}{2}^+,$ and
$\frac{9}{2}^+$)
in the region of 2 MeV above the $n$+${}^{16}$C threshold and a second 
group in the region of 4 MeV above that threshold.  There is also a higher
excited $\frac{1}{2}^+$ state found with both calculations above 6 MeV,
notable by being very broad (half-width 
$\sim 5.6$ MeV with MCAS). The third $\frac{3}{2}^+$
state in the MCAS spectrum (centroid $E_x \sim$ 4.5 MeV) is
also very broad (half-width $\sim$ 4 MeV).
Our MCAS result shows more states, including two very narrow ones
of spin-parity $\frac{13}{2}^+$ and $\frac{11}{2}^+$ at 
3.82 and 4.16 MeV excitation respectively.

\begin{table}[h]
\begin{ruledtabular}
\caption{\label{C17table} Spectra for ${}^{17}$C to 6 MeV excitation. 
Units are MeV.
The $n$+${}^{16}$C threshold lies at $0.728$~MeV.}
\begin{tabular}{ccccccc}
 & Ref.~\cite{Bo07} & & Ref.~\cite{Ra96} & & MCAS & \\
$J^\pi$ & $E$ & $\frac{1}{2}\Gamma$ & $E$ & $J^\pi$ & $E$ & 
$\frac{1}{2}\Gamma$ \\
\hline
$\frac{3}{2}^+$ &  0.00 & & 0.000 & $\frac{3}{2}^+$ & 0.00 & \\
$\frac{1}{2}^+$ & 0.21 & & 0.292 & $\frac{1}{2}^+$ & 0.201 & \\
$\frac{5}{2}^+$ & 0.31 & & 0.295 & $\frac{5}{2}^+$ & 0.390 & \\
\hline
 & & & (1.18) & $\frac{7}{2}^+$ & 1.196 & 2x10${}^{-7}$ \\
$(\frac{3}{2}^+, \frac{7}{2}^+)$ 
& 2.06 & 0.50 & (2.25) & $\frac{9}{2}^+$ & 2.409 & 5x10$^{-5}$\\
 & & & (2.64) & $\frac{5}{2}^+$ & 2.714 & 0.017 \\ 
$\frac{9}{2}^+$ & 3.10 & 0.20 & & $\frac{3}{2}^+$ & 3.120 & 0.100\\
 & & &        & $\frac{13}{2}^+$ & 4.018 & 5x10$^{-5}$ \\
 & & & (3.82) & $\frac{7}{2}^+$ & 4.038 & 0.086 \\
$(\frac{5}{2}^+, \frac{7}{2}^+, \frac{9}{2}^+)$ 
& 4.25 & 0.28  & & $\frac{11}{2}^+$ & 4.261 & 2x10$^{-5}$ \\
 & & & & $\frac{7}{2}^+$ & 4.843 & 0.008 \\
 & & & & $\frac{3}{2}^+$ & 5.063 & 2.075\\
 & & & & $\frac{9}{2}^+$ & 5.235 & 0.044 \\
 & & & & $\frac{5}{2}^+$ & 5.300 & 0.172 \\ 
\end{tabular}
\end{ruledtabular}
\end{table}
Specifics of the states in the ${}^{17}$C spectrum are listed
in Table~\ref{C17table}. Columns 1 to 3 display values of
spin-parity $J^\pi$, excitation energy (or centroid) $E$, and half-widths 
$\frac{1}{2} \Gamma$, of resonances ascertained in a study~\cite{Bo07} 
of three neutron transfer cross sections for ${}^{12}$C scattering from 
${}^{14}$C.
The excitation energies of states shown in Fig.~5 of Ref.~\cite{Ra96} 
are listed in column 4, while the MCAS results are displayed in 
columns 5, 6, and 7. The three nucleon transfer reaction widths in
general  do not match those from the MCAS evaluation but the two
sets are quite different; the latter being single nucleon removal 
values.

The resonant states found using MCAS in the region of 3 MeV 
excitation have widths that agree well with most of the 
matching ones from the MCM evaluation~\cite{Ti10}. 
Widths quoted in Ref.~\cite{Ti10} are $\left. \frac{7}{2}^+\right|_1$
($10^{-12}$ MeV), $\left. \frac{9}{2}^+\right|_1$ ($10^{-6}$ MeV), and
$\left. \frac{5}{2}^+\right|_2$ ($0.015$~MeV). 

We note also that Satou {\it et al.}~\cite{Sa08} from their measurements
of 70$A$ MeV ${}^{17}$C radioactive ion beam scattering from hydrogen
suggest that there are resonances in ${}^{17}$C at 2.2, 3.05, and 6.13
MeV with spin-parities anticipated to be $\frac{7}{2}^+, \frac{9}{2}^+$,
and $\frac{5}{2}^+$ respectively. In a recent article~\cite{Fo11},
results of using a simple shell model approach and considering
single-particle widths suggests that the 2.2 MeV resonance is
in fact two or three narrow but closely spaced ones.

\begin{figure}[h]
\scalebox{0.8}{\includegraphics*{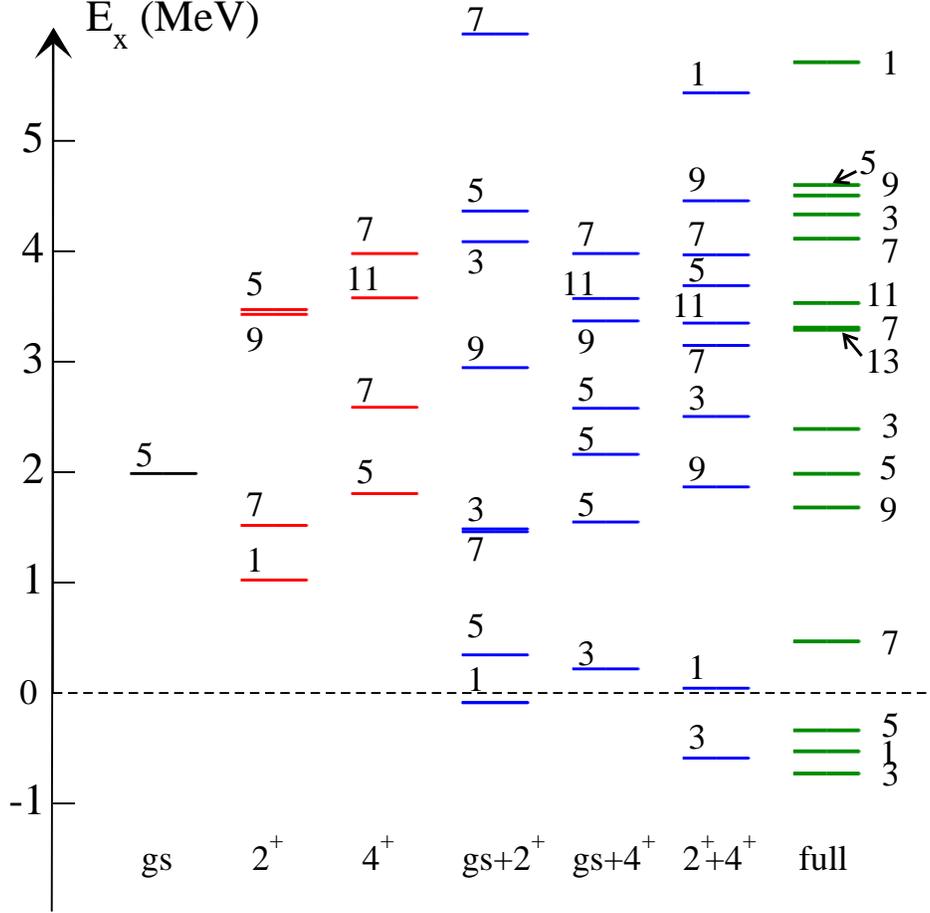}}
\caption{\label{components}
Spectra from MCAS evaluations restricting couplings to
single and  two channels of the three compared
with the full three-channel spectrum. Again the energy 
scale is set with the $n$+${}^{16}$C threshold as zero
to emphasise the sub-threshold from resonance states in
each evaluation.}
\end{figure}

${}^{17}$C has been noted~\cite{Mu09,Ti10} as having a ``peculiar"
structure which may be 
connected with the neutron separation energy from the ground 
state being only $0.728$~MeV. That is typical of a halo nucleus. Indeed, 
the channel-coupling interaction we require to give the spectrum of
${}^{17}$C reflects that with diffuseness being large. Also,
features of this interaction resemble those required~\cite{Ti10}
in a two-body potential model of this system,
{\it viz. ``The bound ${}^{17}${\rm C} spectrum cannot be understood
in the two-body potential model with deformation and the 2$^{\rm +}$ 
excitation of the ${}^{16}${\rm C} core either, if standard sets of 
potentials are used.  An $\ell$-dependent and nonstandard 
spin-orbit $n${\rm +}${}^{16}${\rm C} potential must be used for these
purposes.''}  That is illustrated in Fig.~\ref{components}
which shows how spin states arise from underlying
components of the coupled-channel approach.  Calculations have been made
for each subdivide of the full three target state coupled-channel
problem.
The numerals indicate two times the spin values.
The first three columns from the left give the states found when each
state alone is considered as a single channel problem. Obviously, within
the searched energy range (to 6 MeV), the nucleon on the ground state  
gives only a single state from adding a $d_{\frac{5}{2}}$ neutron. The
addition of a neutron (probably into a $d_{\frac{5}{2}}$ single-particle 
state) gives the set shown for the single channels of the $2^+$
and $4^+$.  The order shown is due to the spin and angle
dependent features used in the base interaction. The existence of the
second $\frac{7}{2}^+$ state when a neutron is added to the isolated 
$4^+$ state may be reflecting addition in a $d_\frac{3}{2}$ state. 

The next three columns are the spectra found when two of the three
${}^{16}$C states are allowed in the coupling. The results found
coupling the ground and $2^+$ states are very similar to what was
published in an earlier paper~\cite{Ka08}; differences reflecting
changed interaction parameter values. The inclusion of the $4^+$
state, whether in a 2- or the full 3-state coupling study, leads 
to a richer spectrum and  changes the order of states.
Of note, only evaluations in which the $2^+$ state is
included give a low lying $\frac{1}{2}^+$ state. Moreover, the three-state 
coupling markedly moves the energy values from that found
otherwise.

The key feature is finding the three closely spaced sub-threshold
states and in this spin order. Non-trivial changes of the parameters,
notably with the deformations, radius and diffuseness to smaller
values, cause such packing to be lost.
The coupling of the $4^+$ state in ${}^{16}$C with the ground
 and $2^+$ states causes notable changes to the predicted spectrum;
changes that better align with (so far) experimentally defined states
in ${}^{17}$C, a few of which have been assigned spins.  Clearly
we predict many other resonance states, but whether they exist or
can be found if they do, remains an open question. It is most unlikely
that any direct $n$+${}^{16}$C experiment will ever be done so
one must rely on some surrogate approach, such as ${}^{16}$C$(d,p)$,
or some study that identifies neutron emissions from ${}^{17}$C.

\section{MCAS evaluations of ${}^{17}$Na as a
$p$+${}^{16}$Ne system}
\label{mcasNa}

${}^{17}$Na ($p$+${}^{16}$Ne) is the mirror system to ${}^{17}$C 
($n$+${}^{16}$C) but, to date, none of its states have been
identified.  Its ground state mass is not listed in the Ame2003 mass
table~\cite{Au03}. However, it is expected~\cite{Ti10} that 
the spectrum of this 
nucleus should have a number of low-excitation resonant states,
some of which are to be quite narrow. The MCM study
of the structure anticipates that the unbound ground state
should lie around 2.4 MeV above the proton-${}^{16}$Ne threshold;
a value notably less than 3.65 MeV expected using the  Kelson and Garvey
formula~\cite{Ke66}.

The MCAS calculations of the $p$+${}^{16}$Ne system were made
adding a Coulomb potential to 
the nuclear interaction set by the analysis of the ${}^{17}$C spectrum.
The Coulomb potentials are those derived from a Woods-Saxon charge
distribution having the same parameters (geometry and deformation) as the 
nuclear interaction.
The spectrum of ${}^{16}$Ne known to date is just the ground state ($0^+$),
and an excited $2^+$ state at 1.69 MeV.  We presume, based upon the 
spectrum of its mirror, ${}^{16}$C,  that there will be
a $4^+$ state in the vicinity of 4 MeV excitation. 

\subsection{Sharp target state results}

\begin{figure}[h]
\scalebox{0.7}{\includegraphics*{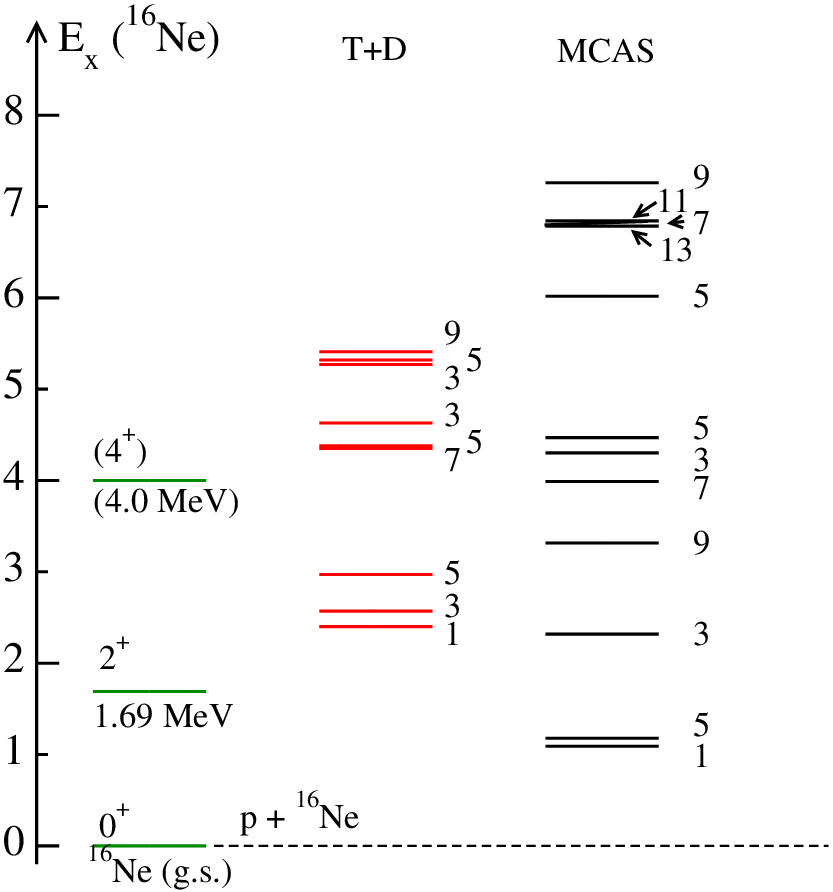}}
\caption{\label{Na17spect}
Results from model evaluations of the spectrum of ${}^{17}$Na.
The energy scale is relative to the $p$+${}^{16}$Ne threshold.}
\end{figure}

Initially we suppose that the three states in ${}^{16}$Ne 
are like those assumed for ${}^{16}$C namely to have zero widths. 
Using those states in an MCAS evaluation 
with a Coulomb interaction added to the nuclear one
gave the spectra of
${}^{17}$Na ($p$+${}^{16}$Ne) as indicated
in Fig.~\ref{Na17spect}. 
The resulting spectrum (labelled MCAS) is compared with that given in 
Ref.~\cite{Ti10} (identified in the figure by the label T+D).  
The three states of 
${}^{16}$Ne used in both studies are shown on the left of this diagram.
The spins of the (positive parity) states found for ${}^{17}$Na
are indicated by two times  their values and are shown for each level
found.  
The excitation spectra depicted for  ${}^{17}$Na 
have the order of the spins interchanged and 
the size of energy gaps differ from those in ${}^{17}$C.
The three lowest excited states from each calculation have spin-parities of
$\frac{1}{2}^+, \frac{3}{2}^+,$ and $\frac{5}{2}^+$. But while the
$\frac{3}{2}^+$ state is the first 
excited state in the MCM spectrum, it 
is the second excited state the MCAS evaluation.
The energy gaps of the three lowest states predicted by each
calculation are different. 

It may be thought that, the Thomas-Ehrman (TE) shift could make the (MCAS)
$\frac{1}{2}^+$ state, in particular,  change noticeably in energy.
The TE shift is of note when there is a weakly bound $s$-wave proton
coupling to a core state.
However, since the dominant attributes
in the low lying states of ${}^{17}$Na in the MCAS evaluation are so 
strongly defined by
coupling of a $d_\frac{5}{2}$ proton to the $2^+$ and $4^+$ 
states in ${}^{16}$Ne, a large TE shift is unlikely since 
the centrifugal barrier helps contain the $d$-wave function. 
\begin{table}[h]
\begin{ruledtabular}
\caption{\label{Na17-spect}
MCAS energies and half-widths of ${}^{17}$Na states.}
\begin{tabular}{cccccc}
 state &  \multicolumn{2}{c}{T+D~\cite{Ti10}} &
\multicolumn{3}{c}{MCAS} \\ 
& $E$(MeV) & $\frac{1}{2} \Gamma$(MeV)$^\star$ &
 $E$(MeV) & $\frac{1}{2} \Gamma$(MeV) & 
$\Bigr. \frac{1}{2} \Gamma\bigl\vert_2$(MeV)\\
\hline
$\frac{1}{2}^+$ & 2.40 & 1.360 & 1.03 & 0.005 & 0.110\\ 
$\frac{3}{2}^+$ & 2.57 & 0.025 & 2.26 & 0.004 & 0.231\\
$\frac{5}{2}^+$ & 2.97 & 0.144 & 1.13 & 0.002 & 0.092\\
$\frac{7}{2}^+$ & 4.35 & 0.025 & 3.98 & 0.036 & 0.242\\
$\frac{5}{2}^+$ & 4.38 & 1.593 & 4.45 & 0.208 & 0.129\\
$\frac{9}{2}^+$ & 5.41 & 0.210 & 3.36 & 0.011 & 0.158\\
$\frac{3}{2}^+$ & 5.27 & 1.383 & 4.28 & 0.103 & 0.249\\
$\frac{5}{2}^+$ & ---  &  ---  & 6.04 & 0.186 & 0.242\\
$\frac{13}{2}^+$ & ---  &  --- & 6.64 & 0.084 & 0.600\\
$\frac{7}{2}^+$ & ---  &  ---  & 6.81 & 0.457 & 0.184\\
$\frac{11}{2}^+$ & ---  &  --- & 6.89 & 0.115 & 0.672\\
\end{tabular}
\end{ruledtabular}

$^\star$ The widths are the summed widths in Table II of
Ref.~\cite{Ti10}. 
\end{table}

With both  model calculations, the states found are 
resonances.  The resonance centroids and half-widths for proton decay 
that the models give 
are listed in Table~\ref{Na17-spect}. 
The widths found with the MCM and MCAS
model are a mix of narrow and
broad while the MCAS results range between 2 and 500 keV.
However, it must be remembered that
${}^{16}$Ne is itself a proton emitter, and so the three 
target states considered should also be treated as resonances.
Doing so with MCAS can have marked effect, in particular upon 
the evaluated compound nucleus widths~\cite{Fr08}.
With the exception of the ground state, widths of the states
in ${}^{16}$Ne are not known, and indeed  even the existence of
a $4^+$ state at $\sim 4$ MeV excitation is not known.  

\subsection{Resonant target state results}

Since the ground state of  ${}^{17}$Na lies above the 
$1p$-, the $2p$-, and the $3p$-emission
thresholds, break-up into those channels give widths
to the (resonant) states of ${}^{17}$Na. 
Mukha {\it et al.}~\cite{Mu10} found the ground state
of ${}^{16}$Ne to have a width of 122 keV while they assigned
a value of 200 keV to the first excited $2^+$ state near
1.69 MeV excitation.  They did not observe a $4^+$ resonance
with centroid $\sim 4$ MeV. They did observe other resonance
states with the lowest in excitation centred about 7.6 MeV.

In MCAS calculations made using resonant states for the target,
we retained the 3-state model taking the ground state to have a width of 122
keV, the $2^+$ (1.69 MeV) state to have a width of 200 keV, and assumed
the existence of a $4^+$ state analogous to that in ${}^{16}$C
centred at 4 MeV and with a width of 1 MeV.
As with recent MCAS studies of other systems~\cite{Fr11,Ca11},
the results of this calculation made only small changes to the centroid
energies of the states in ${}^{17}$Na found using zero
widths for the three ${}^{16}$Ne states. But it had a serious effect upon the
half-widths of the compound nucleus. The  results obtained when the
target states are resonances are listed in column identified
as $\frac{1}{2}\Gamma\bigl\vert_2$ in Table~\ref{Na17-spect}. Comparison
with the values found using zero target-state widths (given in the
adjacent column) show the changes. With the exception of the
$\bigr.\frac{5}{2}\bigl|_2$ and of the $\bigr.\frac{7}{2}\bigl|_2$
which have smaller half-widths, all other resonances increase 
in half-widths; some even more than 100-fold.
The maximum value in the set is 672 keV whereas values of up to 1.6
MeV were found in the MCM evaluation~\cite{Ti10}.

It is essential that new experimental information be provided
for more detailed investigations of model structures for these exotic nuclei.
Nonetheless, a most important theoretical aspect is channel
coupling; the effects of which we discuss next.

\section{Effects of channel coupling in the MCAS approach}

\begin{figure}[h]
\scalebox{0.8}{\includegraphics*{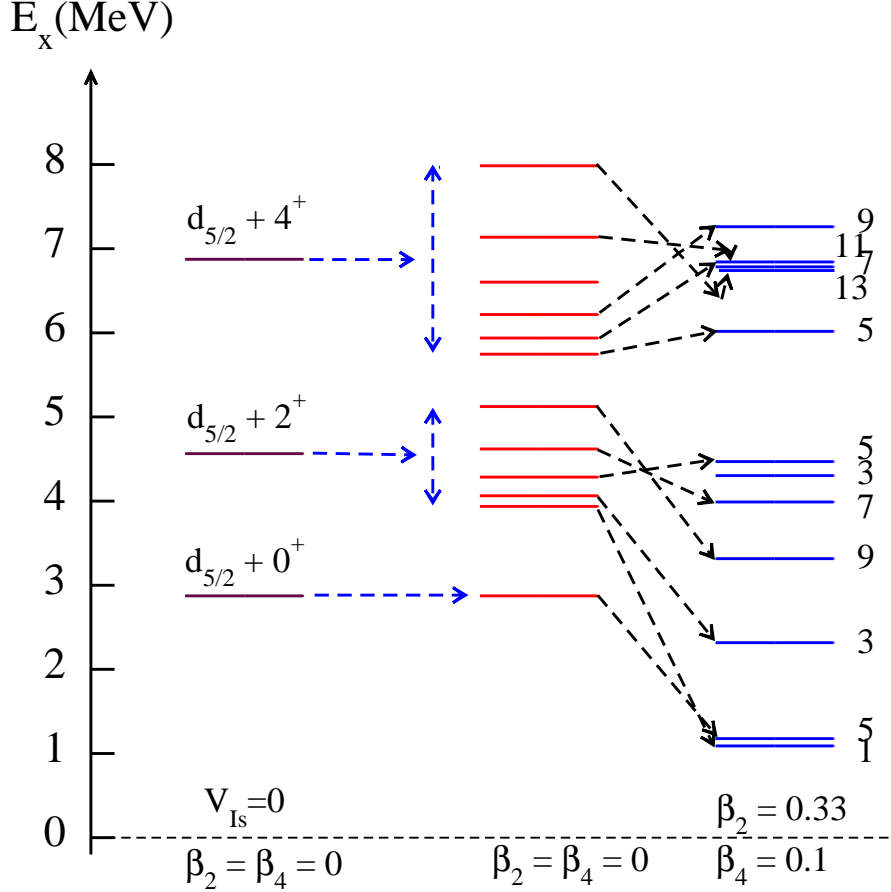}}
\caption{\label{Zero-def}
MCAS spectra for ${}^{17}$Na  found with no deformation and 
$V_{Is} = 0$ (left), with no deformation (center), and the 
full result (right).}
\end{figure}

In Fig.~\ref{Zero-def} we show  MCAS spectra found 
with (on the right)  and without (in the center) deformations of the 
nuclear interactions. The spectrum on the left is that when both the
deformation parameters and the spin-spin strength, $V_{Is}$, are set to
zero. 
As usual, the spins of the states considered
are indicated by two times
their values and matching pairs (of spin values) are indicated by the 
dashed lines linking the no deformation to the full results.
Of course the deformations result in states that are admixtures of 
those of the same spin-parity shown in the center spectrum.
setting the spin-spin interaction to zero and having no deformation
shows that
these states are the result of coupling a $1d_{\frac{5}{2}}$ 
proton to the
three selected states for ${}^{16}$Ne. Such coupling allows twelve
states as a basis, namely the set $\frac{1}{2}^+$, two $\frac{3}{2}^+$,
three $\frac{5}{2}^+$, two $\frac{7}{2}^+$, two $\frac{9}{2}^+$,
an $\frac{11}{2}^+$, and a $\frac{13}{2}^+$. 
However, with larger (quadrupole) deformation, extra states
built largely upon the coupling of $2s_{\frac{1}{2}}$ and/or
$1d_{\frac{3}{2}}$ proton states to the three ${}^{16}$Ne ones
can be found in the spectrum within that excitation energy range.
Clearly the deformation of the nuclear interaction has a very
large effect on the resultant spectrum.

\begin{figure}[h]
\scalebox{0.8}{\includegraphics*{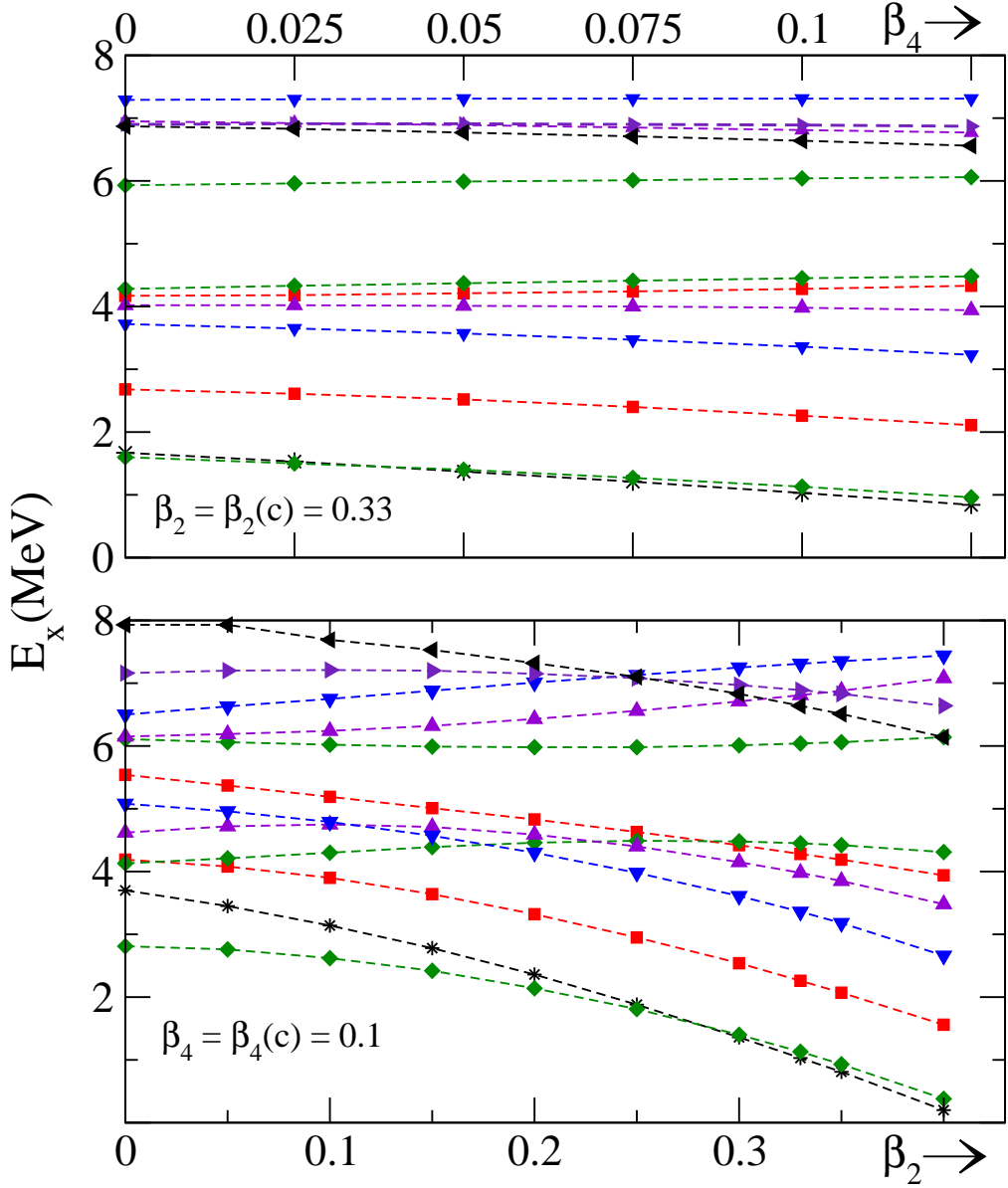}}
\caption{\label{Beta-var}
Variations of the spectrum of ${}^{17}$Na upon changing the
quadrupole (bottom) and hexadecapole (top) deformations.
The $\frac{1}{2}^+$ state is depicted by the stars while $\frac{3}{2}^+,
\frac{5}{2}^+, \frac{7}{2}^+$ and $\frac{9}{2}^+$ states are identified by
the filled squares, diamonds, up-triangles, and down-triangles
respectively. The high spin states, $\frac{11}{2}^+$ and
$\frac{13}{2}^+$, are portrayed by the left- and right-triangles
respectively.}
\end{figure}

In Fig.~\ref{Beta-var} are shown the changes in the spectrum
that occur as the quadrupole (bottom segment) and hexadecapole (top 
segment) deformations are varied separately. 
The energies are values relative to the $p$+${}^{16}$Ne threshold.
The set of states at $\sim 8$~MeV excitation are hardly influenced
by the hexadecapole deformation, but, albeit small, that deformation 
causes some changes to the lower excitation states. The  three
lowest excitation states, of spins $\frac{1}{2}^+, \frac{5}{2}^+$ and 
$\frac{3}{2}^+$ in sequence, become 
more bound with increasing $\beta_4$,
and by about the same amount.
Likewise the second band of resonant states show an interesting variation
with the conjectured $\frac{3}{2}^+$ and
$\frac{5}{2}^+$ states
becoming less bound with increasing hexadecapole deformation, while the 
$\frac{7}{2}^+$ state is little changed but the $\frac{9}{2}^+$
state becomes noticeably more bound.  

The variation in the spectrum caused by changes to the size of
quadrupole deformation is shown in the bottom segment of 
the top panel of this figure. With increasing quadrupole
deformation, the states of the spectrum spread with most, but not
all, decreasing in excitation energy. The influence on the
lowest three (resonant) states is most marked for changes in
$\beta_2$ in the range 0.3 to 0.4, with the $\frac{1}{2}^+$
state eventually becoming the ground state in the MCAS model
evaluations.

\section{Conclusions} 

We have used the MCAS approach to suggest a low-excitation spectrum 
for the exotic nucleus ${}^{17}$Na treated as a $p$+${}^{16}$Ne system.
The mirror isospin concept was used to define 
the nuclear $p$+${}^{16}$Ne
interaction as that which gave a reasonable low excitation spectrum 
for ${}^{17}$C when treated as a $n$+${}^{16}$C coupled-channel problem.
Three states, a $0^+$ (ground) a first excited $2^+$ ($\sim 1.7$ MeV),
and a $4^+$ ($\sim 4$ MeV), were taken as the target states in the
coupled-channel studies.  An interaction was found that gave
sub-threshold and low excitation resonances for ${}^{17}$C with centroid
energies in quite reasonable agreement with the limited set of known
values. The ${}^{17}$C spectrum contains both sharp and narrow resonances.
By adding Coulomb interactions for the interaction between a proton
and the individual states of ${}^{16}$Ne, the spectrum of the nucleon
emissive ${}^{17}$Na also contained narrow and broad resonances.
Those MCAS results had some similarities with, but also a
number of variations from, the spectrum of ${}^{17}$Na defined by the
MCM of the $p$+${}^{16}$Ne system~\cite{Ti10}.
Both evaluations gave a resonance state spectrum for that nucleus
which is known to lie outside of the proton drip line; spectra that
had both narrow and broad resonances.  The MCAS result placed the 
resonant ground state of ${}^{17}$Na an MeV or so lower than that of
the MCM study and reversed the spins of the first and second excited
states. 
A larger energy ($\sim 3$ MeV) 
is expected from an older estimate~\cite{Ke66}
though the actual mass excesses of relevant nuclei 
that calculation were not well fixed. 
Nonetheless, our results confirm the general
findings of the MCM study~\cite{Ti10}; notably that channel coupling effects 
are crucial
in ascertaining both the centroid energies as well as a range of widths
from narrow to broad in low-excitation spectra. Such channel couplings were
also important in
predicting the character of another nucleus that lies just beyond the
proton drip line, namely ${}^{15}$F~\cite{Ca06a}.  
In detail, our results and those of the MCM study do not match well
with the effects of using actual resonance states to describe ${}^{16}$Ne
being very significant as to what half-widths are found for the
resonances in ${}^{17}$Na.

Of course, just what couplings are important,
how many channels are relevant, what are the optimal nuclear and Coulomb 
interactions, and what microscopy underlies the spectrum, await improved
experiments of high quality from which more details of the nuclei,
${}^{16,17}$C, ${}^{16}$Ne, and ${}^{17}$Na, can be ascertained.

\section*{Acknowledgments}
SK acknowledges support from the National Research Foundation of South
Africa. 
JPS acknowledges support from the National Sciences and Engineering 
Research Council (NSERC) of Canada.

\bibliography{Mass17}

\end{document}